\documentclass{ws-procs975x65}

\usepackage{amsmath}
\usepackage{amsfonts}
\usepackage{amssymb}
\usepackage{dsfont}
\usepackage{latexsym}

\begin{document}

\title{The electro-weak model as a phenomenological issue of multidimensions}

\author{Francesco Cianfrani$^*$ and Giovanni Montani$^{*\dag}$}

\address{$^*$ICRA---International Center for Relativistic Astrophysics\\
Dipartimento di Fisica (G9), Universit\`a  di Roma, ``La Sapienza",\\
Piazzale Aldo Moro 5, 00185 Rome, Italy.}

\address{$^\dag$ENEA-C.R. Frascati (U.T.S. Fusione),\\ via Enrico Fermi 45, 00044 Frascati, Rome, Italy.}

\address{E-mail: francesco.cianfrani@icra.it\\ montani@icra.it}

\begin{abstract}
We consider Kaluza-Klein theories as candidates for the unification of gravity and the electro-weak model. In particular, we fix how to reproduce geometrically the interaction between fermions and gauge bosons, in the low energy limit.\end{abstract}


\bodymatter

\section*{}
One of the main task of modern theoretical physics is an explanation for the different kind of interactions in nature. Among them, gravity plays a special role, being described by a classical and geometrical theory (General Relativity); therefore, the most difficult obstacle seems to be its unification with other forces. While it is a common opinion that in a quantum formulation the gravitational field could look more similar to gauge ones, no evidence that Quantum Gravity will imply a unification exists (for example, in Loop Quantum Gravity no unification happens \cite{rov}). Furthermore, a different perspective exists, that is to move the formulation of fundamental interactions close to that for gravity, by a geometrical interpretation for gauge fields degrees of freedom. A possibility is to arrange bosons into the metric tensor; it implies space-time to have more than four-dimensions. The additional space must be compactified to distances not yet accessible to experiments. Such kind of models are known as Kaluza-Klein theories \cite{acf,cmm}.\\    
In these models, the main assumption is the following form for the metric tensor

\begin{equation*}
\label{c1}
j_{AB}=\left(\begin{array}{c|c}g_{\mu\nu}(x^{\rho})+\gamma_{mn}(x^{\rho};y^{r})\xi^{m}_{\bar{M}}(y^{r})\xi^{n}_{\bar{N}}(y^{r})A^{\bar{M}}_{\mu}(x^{\rho})A^{\bar{N}}_{\nu}(x^{\rho})
& \gamma_{mn}(x^{\rho};y^{r})\xi^{m}_{\bar{M}}(y^{r})A^{\bar{M}}_{\mu}(x^{\rho}) \\\\
\hline\\
\gamma_{mn}(x^{\rho};y^{r})\xi^{n}_{\bar{N}}(y^{r})A^{\bar{N}}_{\nu}(x^{\rho})
& \gamma_{mn}(x^{\rho};y^{r})\end{array}\right)
\end{equation*}

being $x^\mu$ and $g_{\mu\nu}$ the four-dimensional coordinates and metric, respectively, while $y^m$ and $\gamma_{mn}$ the analogous ones on the extra-dimensional space, endowed with Killing vectors $\xi^{m}_{\bar{M}}$.\\ 
The possibility to geometrize a gauge theory is encoded in the existence of a homogeneous manifold, whose isometries reproduce the algebra of the gauge group in the following way
\begin{equation}
\label{a1} \xi^{n}_{\bar{N}}\frac{\partial
\xi_{\bar{M}}^{m}}{\partial y^{n}}-\xi^{n}_{\bar{M}}\frac{\partial
\xi_{\bar{N}}^{m}}{\partial
y^{n}}=C^{\bar{P}}_{\bar{N}\bar{M}}\xi^{m}_{\bar{P}}
\end{equation}

with $C^{\bar{P}}_{\bar{N}\bar{M}}$ structure constant of the Lie group.\\
It is possible to interpret $A^{\bar{M}}_{\mu}$ as gauge bosons since, by the dimensional reduction of the Einstein-Hilbert action, the Yang-Mills Lagrangian density comes out.\\ 
However, the geometrization to be complete requires that fermions have to be introduced in such a scheme and at this point Kaluza-Klein models failed. In particular, it was not satisfied the attempt to reproduce the spectrum of Standard Model fermions starting from multidimensional spinors, by identifying their gauge properties with extra-dimensional degrees of freedom \cite{bl}.\\ 
To overcome some of these difficulties, here we propose a phenomenological approach: 4-fields arise after an average of the equations of motion on the extra-dimensional space, because additional dimensions are undetectable \cite{cm}; as a consequence, geometrical properties related to them are seen in a non-trivial way in a 4-dimensional perspective.\\ 
To geometrize the $SU(2)$ sector of the Electro-Weak model, we consider spinor on $S^3$. According with our approach, low-energy particles are solutions of the Dirac equation averaged on the full space, with a unit weight because of its homogeneity.\\ 
Let us consider the following spinor           
\begin{eqnarray}
\Psi_{r}(x;y)=\chi_{rs}(y)\psi_s(x)=\frac{1}{\sqrt{V}}e^{-\frac{i}{2}\sigma_{(p)rs}\lambda^{(p)}_{(q)}\Theta^{(q)}(y^{m})}\psi_s(x)\label{sp}
\end{eqnarray}
being $V$ the volume of $S^{3}$, $\sigma_{(p)}$ Pauli matrices and $\lambda$ a constant matrix satisfying 
\begin{eqnarray}
(\lambda^{-1})^{(p)}_{(q)}=\frac{1}{V}\int_{S^{3}}
\sqrt{-\gamma}e^{m}_{(q)}\partial_{m}\Theta^{(p)}d^{3}y.
\end{eqnarray}
while for functions $\Theta$ we take
\begin{eqnarray}
\label{theta}\Theta^{(p)}=\frac{1}{\beta}c^{(p)}e^{-\beta\eta}\qquad\eta>0.
\end{eqnarray}
In this way, $\Psi$ is an approximate solution of the averaged Dirac equation, with corrections of order of $\beta^{-1}$ \cite{cm2}; we take it as a good approximation since in the low-energy limit a weak dependence on extra-coordinates, thus a big value for $\beta$, is expected.\\ 
Moreover, by inserting $\Psi$ into the Dirac action, we achieve the geometrization of $SU(2)$ gauge connections; otherwise, some additional and, in general, gauge violating $\beta^{-1}$ terms come out. Therefore, the price to pay for the geometrization is the appearance of some non-renormalizable interactions; however, it is not a problem, because we are performing a low-energy effective theory and gauge violating terms are supressed by the order parameter governing the low-energy limit.\\
Hence, the Electro-Weak model model can be inferred from the geometry of a space-time manifold $V^4\otimes S^1\otimes S^3$, where by the $S^{1}$ manifold the $U(1)$ sector is reproduced. Since in a 8-dimensional manifold spinors have got 16 components, we suggest to recast a quark generation and a fermionic family into the same geoemtrical object, thus explaining the equality of their number. At the end, we reproduce the full fermionic spectrum of the Standard Model from the following two multidimensional spinors for each fermionic family/quark generation
\begin{eqnarray*} 
\Psi_{L}=\frac{1}{\sqrt{V\alpha'}}\left(\begin{array}{c} \chi\left(\begin{array}{c} e^{in_{uL}\theta}u_{L}\\e^{in_{dL}\theta}d_{L}\end{array}\right)\\\chi\left(\begin{array}{c} e^{in_{\nu L}\theta}\nu_{eL}\\e^{in_{eL}\theta}e_{L}\end{array}\right)\end{array}\right)\qquad\Psi_{R}=\frac{1}{\sqrt{V\alpha'}}\left(\begin{array}{c} \left(\begin{array}{c} e^{in_{uR}\theta}u_{R}\\e^{in_{dR}\theta}d_{R}\end{array}\right)\\\left(\begin{array}{c} e^{in_{\nu R}\theta}\nu_{eR}\\e^{in_{eR}\theta}e_{R}\end{array}\right)\end{array}\right)
\end{eqnarray*}     
where the coefficients $n_r$ are related to particles hypercharges, $Y_{r}=\frac{n_{r}}{6}$.\\
In this scenario, a lower bound for $\beta$ can be obtained from experimental limits for processes that violate the conservation of gauge charges. For example, the upper bound on the cross section for the decay of a neutron into a proton plus a couple neutrino-antineutrino gives $\beta\geq10^{14}$.\\
In order to give masses to particles, we introduce a field whose dimensional reduction gives the Higgs field. By assuming the same dependence on extra-coordinates as for spinors, two troubles of the standard Higgs formulation are overwhelmed:
\begin{enumerate}
{\item the appearance of the standard Kaluza-Klein mass term can explain the fine tuning we have to perform, on the parameter of the potential, to stabilize Higg's mass;}
{\item the possibility to assign opposite hypercharges to the two components of the field allows to reproduce a mass term for neutrinos by a standard Yukawa coupling.}
\end{enumerate}
Finally, gauge violating terms give corrections to bosons masses; in particular, from upper bound on the photon mass we get $\beta>10^{28}$.\\ 

\end{document}